\documentclass[aps,prd,preprint,groupedaddress,floatfix]{revtex4-2}

\usepackage{graphicx}
\graphicspath{ {./images/} }
\usepackage{xcolor}

\begin{document}


\title{Surface state mediated ferromagnetism in Mn$_{0.14}$Bi$_{1.86}$Te$_{3}$ thin films}
https://www.overleaf.com/project/627ad71ae6f50f4b48d0bc4a
\author{Ryan Van Haren}
\email{rvanhare@ucsc.edu}
\author{Toyanath Joshi}
\altaffiliation{Current address: Materials Research Lab, University of Illinois, Urbana, IL 61801}
\author{David Lederman}
\affiliation{Department of Physics, University of California Santa Cruz, Santa Cruz, California 95064}


\date{\today}

\begin{abstract}
A spontaneous ferromagnetic moment can be induced in Bi$_{2}$Te$_{3}$ thin films below a temperature $T\approx 16$~K by the introduction of Mn dopants. We demonstrate that films grown via molecular beam epitaxy with the stoichiometry Mn$_{0.14}$Bi$_{1.86}$Te$_{3}$ maintain the crystal structure of pure Bi$_{2}$Te$_{3}$. The van der Waals nature of inter-layer forces in the Mn$_{0.14}$Bi$_{1.86}$Te$_{3}$ crystal causes lattice mismatch with the underlayer to have a limited effect on the resulting crystal structure, as we demonstrate by thin film growth on tetragonal MgF$_{2}$ (110) and NiF$_{2}$ (110). Electronic transport and magnetic moment measurements show that the ferromagnetic moment of the Mn$_{0.14}$Bi$_{1.86}$Te$_{3}$ thin films is enhanced as the Fermi level moves from the bulk conduction band and towards the bulk band gap, suggesting that electronic surface states play an important role in mediating the ferromagnetic order. Ferromagnetic Mn$_{0.14}$Bi$_{1.86}$Te$_{3}$/antiferromagnetic NiF$_2$ bilayers show evidence that the ferromagnetic moment of the Mn$_{0.14}$Bi$_{1.86}$Te$_{3}$ film is suppressed, suggesting the existence of an interface effect between the two magnetic layers.

\end{abstract}


\maketitle

\section{Introduction}
Topological insulator (TI) materials with intrinsic ferromagnetic ordering are interesting due to their wide variety of novel quantum states. One such magnetic quantum state has been observed via the quantum anomalous Hall effect (QAHE) \cite{chang_experimental_2013,jiang_concurrence_2020}. The QAHE is an analog to the quantum Hall effect that exhibits a quantized Hall resistance and dissipation-less edge states but without the need to apply a large external magnetic field, instead relying on the spontaneous magnetic moment of the system to create the observed effects. A more recently discovered quantum state in a magnetic TI is observed via the topological Hall effect (THE), which is associated with the formation of a skyrmion-like magnetic phase at the surface of the material \cite{jiang_concurrence_2020,liu_dimensional_2017}. The THE has potential technological applications in topological spintronics \cite{fert_magnetic_2017}. The discovery of these states suggests a rich parameter space in which to probe quantum mechanical effects in the magnetic TI, with sample geometry, chemical potential, and magnetic ordering all playing important roles.

There are two primary strategies for inducing a spontaneous moment in the surface states of TI. The first is by proximity to a magnetic insulator, usually through the fabrication of a heterostructure of two different materials with an atomically smooth interface which creates some overlap in the electronic and spin states of the two systems. Inducing magnetism by proximity effects has the advantage of maintaining a high crystal quality of the TI, but is limited by a perceived lack of materials that have magnetic and topological properties and which can also support the growth of the TI-magnetic insulator heterostructure \cite{alegria_large_2014}. 
The second method is by doping the TI with magnetic impurities, usually transition metals such as Cr \cite{chang_experimental_2013,kou_interplay_2013,chang_thin_2013}, Mn \cite{hor_development_2010}, or Fe \cite{jo_crossover_2014}. Although the ferromagnetic moment in these systems only orders at low temperatures and the crystal quality is somewhat degraded by the introduction of magnetic dopants, this method of inducing a magnetic moment in the TI has shown the most success in manifesting the quantum states associated with the QAHE and the THE \cite{chang_experimental_2013,jiang_concurrence_2020}. 

In order to maximize the potential utility of magnetically-doped topological insulators, it is important to understand the mechanism of magnetic ordering and how it is affected by microscopic and macroscopic features. In the case of ferromagnetic Mn-doped Bi$_{2}$Te$_{3}$, which has a ferromagnetic moment pointing along the [0001] crystallographic direction of the crystal (using a hexagonal basis) and a Curie temperature of $T_C\approx 16$~K, the mechanism responsible for the magnetic ordering in these materials is unclear. In Mn-doped TI films, proposed mechanisms include Mn clustering \cite{zhang_interplay_2012}, Van Vleck-type susceptibility \cite{lee_ferromagnetism_2014}, and Ruderman–Kittel–Kasuya–Yosida (RKKY) interactions, either through bulk conduction channels or two dimensional (2D) surface states \cite{liu_dimensional_2017, checkelsky_dirac-fermion-mediated_2012,sessi_signatures_2014}.

To gain insight into the mechanism of ferromagnetic ordering in the Mn-doped Bi$_{2}$Te$_{3}$ system, we present here magnetic, electronic, and structural measurements on a series of Mn$_{0.14}$Bi$_{1.86}$Te$_{3}$ (MBT) thin film crystals. Samples were grown on three different types of insulating substrates: non-magnetic Al$_{2}$O$_{3}$ and MgF$_{2}$, and antiferromagnetic NiF$_{2}$. Our results show that subtle, but significant, differences in electronic and magnetic properties develop even between samples that are nominally identical in their growth conditions. By keeping the Mn concentration, film thickness, and other growth parameters constant throughout a series of thin film growths, we approach the question of the nature of induced magnetism in Mn-doped Bi$_2$Te$_3$ topological insulators from a different angle than previous studies, testing variations in the type of substrate, Hall mobility, and charge carrier density for their influence on the magnetic properties of this system. By doing so we present evidence that the ferromagnetic ordering in this system is, in large part, electronic charge carrier mediated and that the antiferromagnetic NiF$_2$ interface acts to reduce magnetization in the MBT film.

This work also demonstrates advancements in the fabrication novel TI-antiferromagnetic insulator bilayers using crystallographic, magnetic, and electronic characterizations of MBT films grown on epitaxial thin films of the antiferromagnetic insulator, NiF$_{2}$. We present evidence that single phase, (0001)-oriented hexagonal MBT films can be grown on the (110) face of tetragonal NiF$_{2}$ and MgF$_{2}$ substrates despite the significant lattice mismatch and difference in crystal structure. Our measurements show that the MBT films grown on NiF$_{2}$ and MgF$_{2}$ have nearly identical crystallographic properties compared to those grown on hexagonal Al$_{2}$O$_{3}$ (0001). These results suggest the existence of a much wider range of potential TI bilayer constructions with unique proximity effects that may emerge at those interfaces.

\section{Methods}
MBT films were grown via molecular beam epitaxy (MBE) in an ultra-high vacuum (UHV) chamber (base pressure $<10^{-10}$~Torr) by sublimating from separate elemental sources of 99.999\% pure Mn, Bi, and Te. Flux ratios and film thickness was determined by measuring the elemental flux rate with a retractable crystal monitor located at the same position as the sample substrate. Reflection high energy electron diffraction (RHEED) oscillations associated with layer by layer growth were measured and used to calibrate the elemental flux to film thickness ratio. Film thickness was measured using x-ray reflectivity (XRR), an independent measurement which reveals that the film thickness calibration from elemental flux had a random error of about 4\%. Each sample was grown under identical conditions, keeping the Mn/Bi flux percentage at $7 \pm 1 \%$ and film thickness to $13.6 \pm 0.5$ nm, approximately 12 quintuple layers (QL). Mn percentage was calculated by a combination of partial pressure ratios during growth and by x-ray fluorescence measurements after growth. Seven distinct MBT samples were grown in a series of five growths or batches over the span of about five months.

For the MBT films grown on Al$_{2}$O$_{3}$ (0001) and MgF$_{2}$ (110), commercially purchased single-crystal substrates were used. For the MBT/NiF$_{2}$ bilayers, a separate UHV chamber was used to first grow the epitaxial NiF$_{2}$ (110) film on a MgF$_{2}$ (110) substrate to a thickness of approximately 30~nm, before MBT film growth. NiF$_{2}$ MBE growths were performed using thermal sublimation of commercially available NiF$_{2}$ source material as described elsewhere~\cite{shi_exchange_2004}. In situ RHEED patterns of all MBT films show similarly smooth, single phase, thin film crystals.
All substrates were annealed at $T = 300^{\circ}$ C for several hours before MBT growth to prepare a clean surface. After MBT film growth, but before removing the sample from the UHV chamber, a 5~nm thick layer of polycrystalline, insulating, non-magnetic MgF$_{2}$ was deposited at room temperature to protect the surface of the film from oxidation in atmosphere. 

X-ray diffraction (XRD) and XRR measurements were performed using Cu K$_\alpha$ radiation from a Rigaku Smartlab thin film x-ray diffractometer. XRD measurements confirmed that all MBT films have the (0001) hexagonal crystal structure of Bi$_2$Te$_3$. The XRR data were analyzed quantitatively by performing non-linear least squares fits using an optical model with the GenX software package~\cite{Bjorck:aj5091} to obtain layer thickness and interface roughness parameters.  The magnetic moments of the films were measured using a Quantum Design MPMS XL superconducting quantum interference device (SQUID) magnetometer by applying an external magnetic field $\mu_0 H = 0.05$~T and measuring as a function of temperature from $T=100$~K to $T=4.5$~K.  

X-ray fluorescence was measured using using a XR-100CR Si detector from Amptek using a monochromated 8.04 keV x-ray emmission from the Cu K$_\alpha$ radiation source used for XRD and XRR measurements. By comparing elemental fluorescence peaks of Bi and Mn across several samples we calculated a 14\% relative error ($\pm 1\%$ Mn percentage) in the Mn concentration of the films \cite{VanHaren2023}.

Seven different samples from five different growths, of which two were grown on Al$_2$O$_3$ (0001), two on MgF$_2$ (110), and three on NiF$_2$ (110), were made into Hall bars for transport measurements by first developing a Hall bar pattern using photolithography, followed by a wet etch in an aqua regia solution to remove the unwanted film. The finished Hall bars were then adhered to chip carriers using commercially available conducting silver paint. Electrical contacts were made between the Hall bar pads and the chip carrier contacts using conducting silver paint and thin copper wire. The Hall bars were 200~$\mu$m wide and contacts were separated by 500~$\mu$m.

The Hall bar samples were loaded into a Janis 12TM-SVM Super VariTemp liquid helium cooled cryostat and measured in magnetic fields of up to 11~T and temperatures ranging from 300~K to $2$~K.  Electronic measurements were made using DC Keithley sources and meters, using a delta mode measurement method to take voltage measurements at each point with alternating pulses of $\pm 10\ \mu$A current. 

Calculations of the anomalous Hall effect (AHE) and carrier density were done by performing a linear regression fit to the measured Hall resistance at $T=2$~K in the magnetic field range between $\mu_0 H=\pm 3\text{ T}$ and $\mu_0 H=\pm 1\text{ T}$, in order to probe only the AHE saturated regions. The intercept on the Hall voltage axis of this fit was used to determine the magnitude of the AHE while the slope, along with the film thickness, was used in the calculation of carrier density. The Hall mobility $\mu$ was calculated from the measured longitudinal resistivity $\rho_{xx}$ of the device together with the carrier density $n$ determined from the transverse resistivity excluding the AHE, $\rho_{xy}=\mu_0 H/ne$, using $\rho_{xx}=1/ne\mu$, where $e$ is the charge of the electron.

\section{Results and discussion}

\subsection{Crystallographic Characterization}
\begin{figure}
    \centering
    \includegraphics{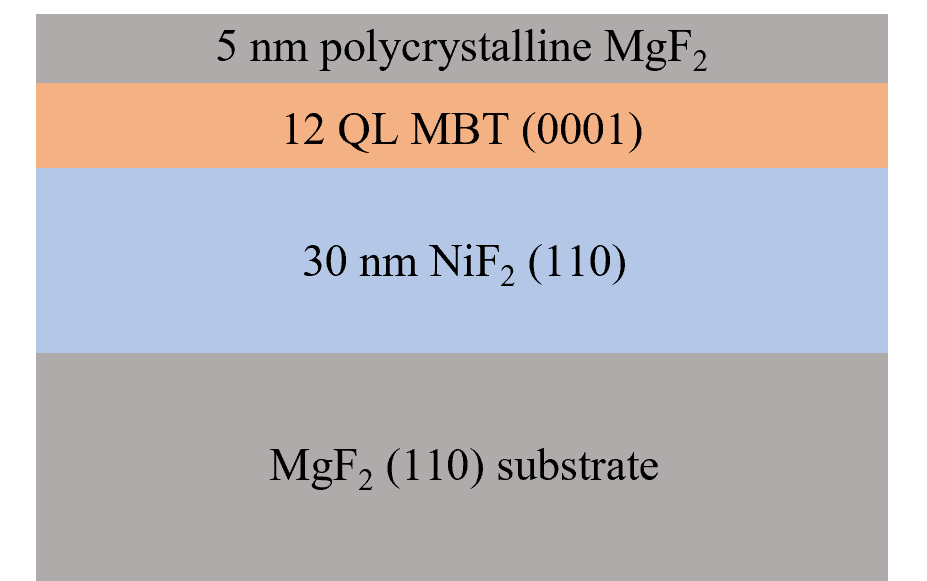}
    \caption{\label{fig:1} Diagram of MBT/NiF$_2$ thin film bilayer sample. Samples grown on MgF$_2$ (110) omit the NiF$_2$ (110) layer, while those grown on Al$_2$O$_3$ (0001) omit the NiF$_2$ (110) layer and have Al$_2$O$_3$ (0001) in place of MgF$_2$ (110).}
    
\end{figure}
\begin{figure}[h]
    \includegraphics{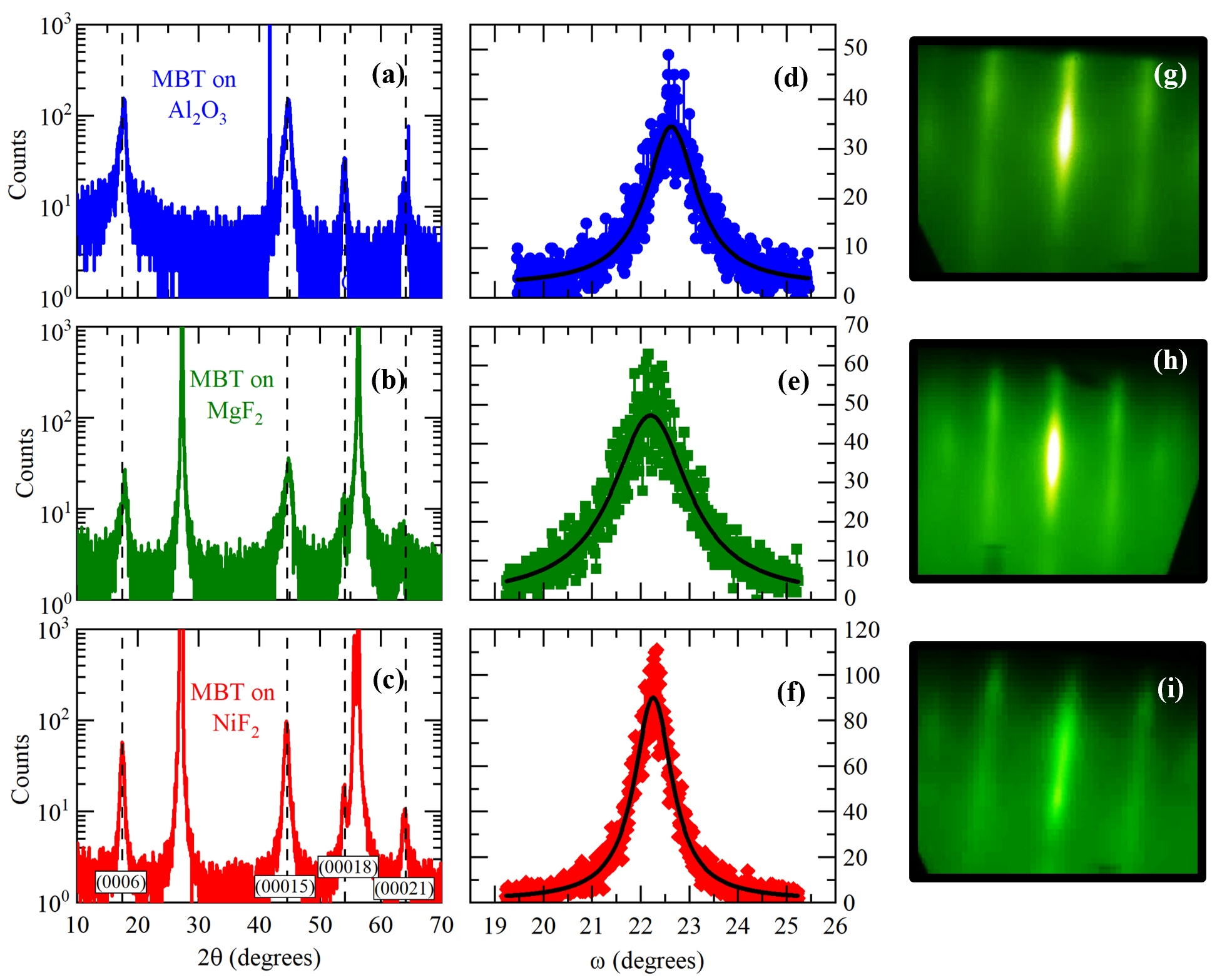}
    \caption{\label{MBTXRD} (a,b,c) XRD pattern of representative MBT films grown on (a) Al$_{2}$O$_{3}$ (0001), (b) MgF$_{2}$ (110), and (c) NiF$_{2}$ (110). Vertical black dashed lines indicate expected location of Bi$_{2}$Te$_{3}$ (000$\ell$) diffraction  peaks. Large, unmarked peaks correspond to the substrate diffraction peaks. (d,e,f) Rocking curve measurements and Voigt line shape fits to the (00015) diffraction peak of the representative films. (g,h,i) RHEED patterns of the corresponding MBT films obtained in situ after film growth but before MgF$_{2}$ capping layer deposition.}
\end{figure}

The MBT thin films were grown on three types of substrates, hexagonal Al$_{2}$O$_{3}$ (0001), tetragonal MgF$_{2}$ (110), and tetragonal NiF$_{2}$ (110). A diagram of the sample structure is given in Fig.~\ref{fig:1}. XRD patterns of three representative MBT films grown on the three different substrates studied are shown in Fig.~\ref{MBTXRD}. The positions of the MBT diffraction peaks show no significant shifts relative to the expected peak positions of pure (000$\ell$) orientation Bi$_{2}$Te$_{3}$, and no evidence of additional peaks that could be associated with other crystal structures or orientations. This result is consistent with a Bi$_2$Te$_3$ film with randomly distributed Mn dopants substituting at the Bi sites, rather than with the layered MnBi$_2$Te$_4$ family of crystals, which has a larger (0001) lattice parameter due to the addition of an ordered Mn layer \cite{yan_crystal_2019}. RHEED patterns of the MBT films taken in-situ are shown in the insets of Fig. \ref{MBTXRD}, and show bright, sharp streaks associated with smooth, single phase growth at the surface. It has been shown previously that attempting to incorporate too much Mn into the thin film crystal will degrade the structure significantly \cite{lee_ferromagnetism_2014}. However, the result presented here is evidence that small amounts of Mn dopants, such as the $7 \pm 1 \% $ doping of Mn used in this study, can be incorporated into the Bi$_{2}$Te$_{3}$ film without significant degradation of the crystal structure. Furthermore, MBT films grown on unconventional, tetragonal MgF$_2$ (110) and NiF$_2$ (110) films (which are themselves grown epitaxially on MgF$_2$ (110) substrates) are of very similar crystal quality to the film grown on Al$_2$O$_3$, as shown in Fig. \ref{MBTXRD}. This conclusion is supported by rocking curve measurements of the MBT (00015) XRD peaks, where the calculated full-width-half-maximum values from Voigt line shape fits are $0.9 \pm 0.3^\circ$, $1.5 \pm 0.5^\circ$, \& $1.8 \pm 0.7^\circ$ for MBT on Al$_2$O$_3$, MgF$_2$ and NiF$_2$, respectively. We believe that the van der Waals bonding between adjacent layers of the MBT crystal makes it relatively insensitive to the lattice of the substrate it is grown on, so long as the substrate surface is sufficiently clean and smooth. 

\begin{figure}[h]
    \includegraphics{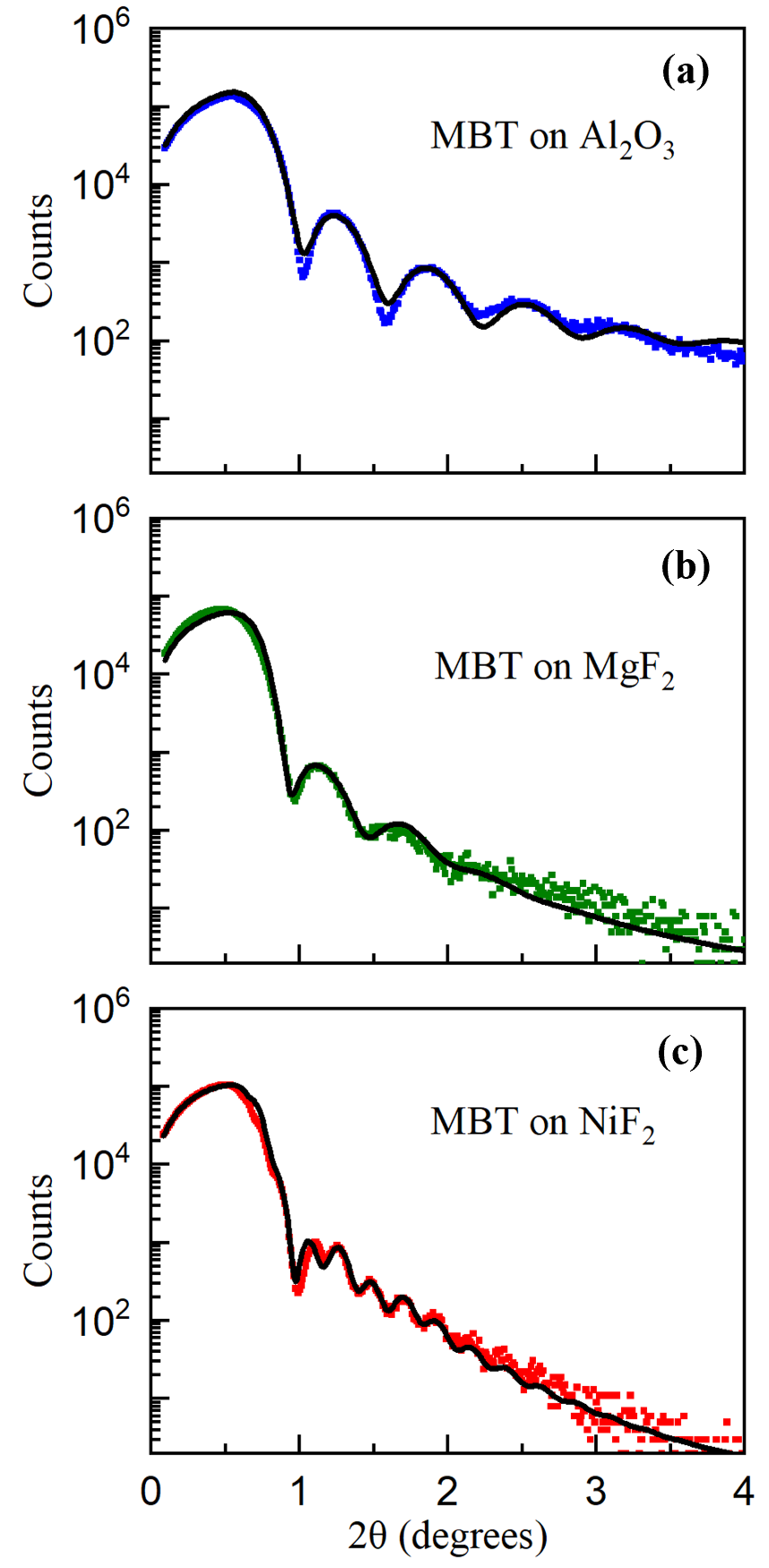}
    \caption{\label{fig:3} XRR pattern of MBT films grown on (a) Al$_{2}$O$_{3}$ (0001), (b) MgF$_{2}$ (110), and (c) MgF$_2$/NiF$_{2}$ (110). Solid black curves indicate fits to the measured data.}
\end{figure}

\begin{table}[h]
    \caption{\label{tab:XRR}Interface roughness ($\sigma$) and film thickness ($t$) parameters extracted from fits to XRR data shown Fig.~\ref{fig:3} in units of nm. ``sub" refers to the substrate, ``cap" refers to the MgF$_2$ capping layer, ``NA" = not applicable.}
\begin{ruledtabular}
\begin{tabular}{cccccccc}
    Substrate & $\sigma_{\text{sub}}$ &$t_{\text{NiF}_2}$ &$\sigma_{\text{NiF}_2}$ & $t_{\text{MBT}}$& $\sigma_{\text{MBT}}$&$t_{\text{cap}}$&$\sigma_{\text{cap}}$\\
    Al$_2$O$_3$ (0001) & 0.7 & NA & NA& 12.7 & 0.1& 4.8 & 2.3\\
    MgF$_2$ (110) & 1.5 & NA &NA & 13.8 & 0.2& 4.5 & 1.7\\
    MgF$_2$/NiF$_2$ (110) & 1.2 & 22.7 & 1.8 & 13.3 & 0.4 & 4.9 & 1.6 \\
\end{tabular}
\end{ruledtabular}
\end{table}

Raw XRR data and the fits to the data using GenX are shown in Fig.~\ref{fig:3}. Table~\ref{tab:XRR} shows the layer thickness and interface roughness values extracted from the fits. The MBT film thickness values are consistent with the expected thickness from calibration of molecular beam flux during growth, corresponding to approximately 12 QL of MBT. It is interesting to note that although the MgF$_{2}$ and NiF$_{2}$ substrates host MBT interfaces that are nearly twice as rough as the MBT interface with Al$_{2}$O$_{3}$, the final surface roughness of the MBT layer is not similarly as rough, likely due to Van der Waals bonding with the substrate, rather than epitaxial growth. These XRR data, in conjunction with XRD and RHEED data, offer compelling evidence that smooth, single phase MBT films can be grown on the tetragonal (110) surfaces of MgF$_{2}$ and NiF$_{2}$.

\subsection{Magnetic Moment Measurements}

\begin{figure}
    \includegraphics{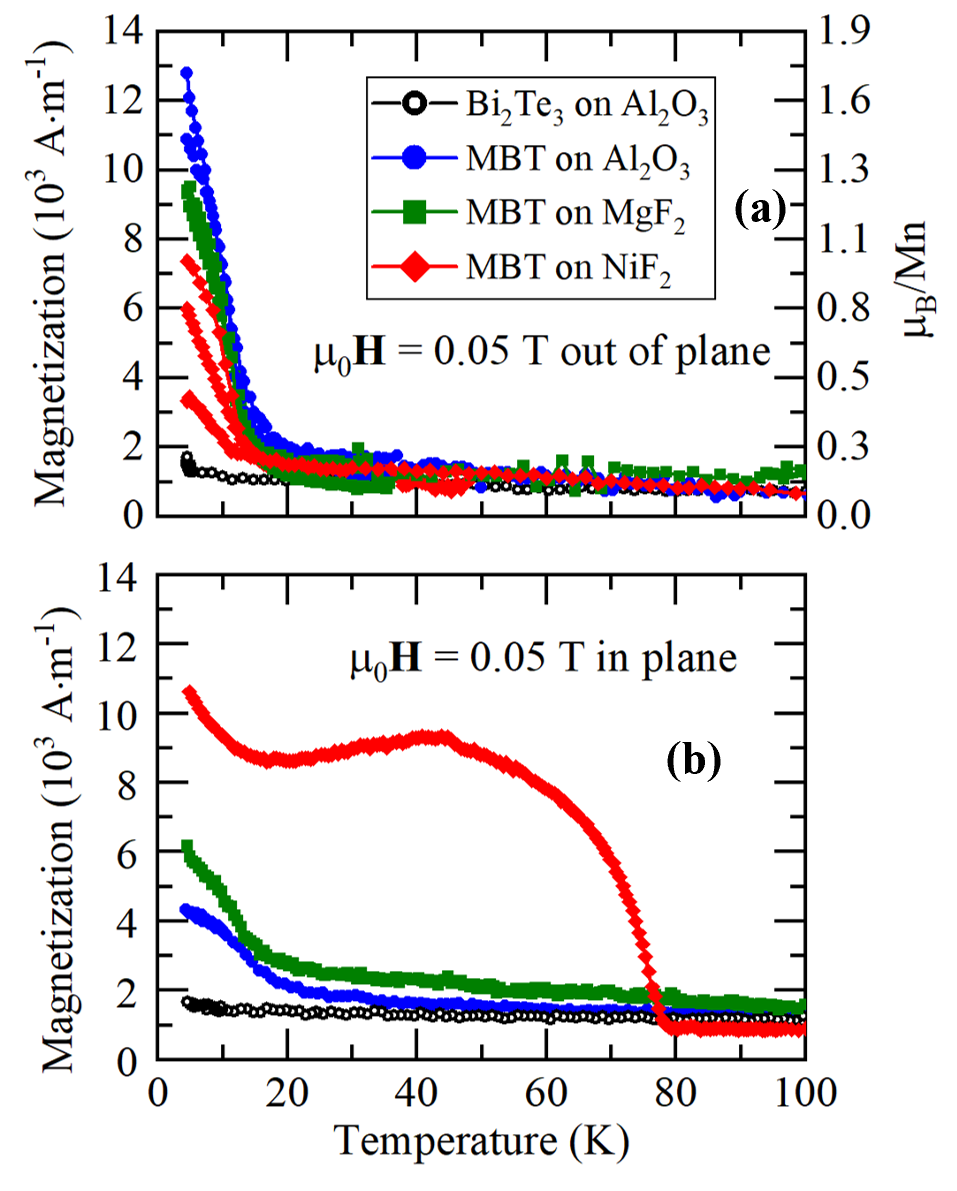}
    \caption{\label{MBT MvT} Field cooled magnetization as a function of temperature, measured along the same direction as the applied field, of a single run from several different MBT films grown on Al$_{2}$O$_{3}$ (shown in blue), MgF$_{2}$ (shown in green), and NiF$_{2}$ (shown in red), and undoped Bi$_{2}$Te$_{3}$ grown on Al$_{2}$O$_{3}$ (shown in black). (a) Magnetization measured measured perpendicular to the film surface, along the [0001] axis of the MBT and Al$_{2}$O$_{3}$ crystals, and the [110] direction of the MgF$_{2}$ and NiF$_{2}$ crystals, with $\mu_0H = 0.05$~T. (b) Magnetization as a function of temperature of three representative MBT samples measured parallel to the film plane, along the [1$\bar{1}$0] direction of the MgF$_{2}$ and NiF$_{2}$ crystals, with $\mu_0H = 0.05$~T.}
\end{figure}

While doping Bi$_{2}$Te$_{3}$ with a small amount of Mn does not significantly disrupt the crystal structure of the Bi$_{2}$Te$_{3}$ film, it does lead to the formation of a spontaneous ferromagnetic moment. Magnetization measurements as a function of temperature of the MBT films are presented in Fig.~\ref{MBT MvT} with an applied field $\mu_0H=0.05$~T. Figure \ref{MBT MvT}(a) shows data from a single run from all seven samples used in this study, with the magnetic moment measured perpendicular to the surface of the film, along the [0001] direction of the MBT film. A clear transition to a ferromagnetic state is observed in all MBT samples at an average temperature of $T_C = 15.5 \pm 1.0$ K, where the error is dominated by small variations in transition temperature between samples \cite{VanHaren2023}. We observed no clear correlation between the transition temperature and type of substrate used. Figure \ref{MBT MvT}(b) shows the magnetic moment when the samples are rotated $90^\circ$ to measure the moment in the plane of the film. In this direction, the ferromagnetic transition of the MBT film is more rounded and suppressed when compared to the out-of-plane direction. This magnetic anisotropy and transition temperature of the MBT film is consistent with previous studies of Mn doped Bi$_{2}$Te$_{3}$, which has a magnetic easy axis along the [0001] crystallographic direction and a transition temperature around 15 K \cite{hor_development_2010,lee_ferromagnetism_2014}. It is important to note here that the magnetization values of the MBT/NiF$_2$ bilayers at $ T = 4.5 $ K are each lower than the magnetization values of any of the MBT films on non-magnetic substrates. This behavior suggests that the NiF$_2$ may be acting to reduce the out-of-plane magnetization of the MBT films. More evidence of this effect will be presented below in measurements of the electronic transport of the films. 

Measuring the magnetization along the in-plane direction reveals the magnetic behavior of the NiF$_2$ layer in the MBT/NiF$_2$ bilayers. While Al$_{2}$O$_{3}$ and MgF$_{2}$ are non magnetic, NiF$_{2}$ is an insulating antiferromagnet with a transition temperature of 73 K in bulk, and a N\'eel vector that orders in the $a\text{-}b$~plane, preferentially along the [100] or [010] axes \cite{moriya_theory_1960,borovik_1973}. NiF$_{2}$ also exhibits weak ferromagnetism due to a Dzyaloshinskii-Moriya interaction that causes a spontaneous canting of the antiferromagnetic moments in the $a\text{-}b$~plane \cite{moriya_theory_1960}. As a result, a ferromagnetic transition is evident in the MBT/NiF$_{2}$ bilayer, as shown by the magnetic response in Fig.~\ref{MBT MvT}(b). The NiF$_{2}$ transition temperature in these thin films is shifted from the expected 73 K bulk value to 78 K due to out-of-plane tensile strain in the NiF$_{2}$ thin film crystal resulting from the epitaxial growth on MgF$_2$ (110). The observed correlation between tensile strain in the [110] direction and transition temperature shown here is in agreement with previous studies of NiF$_{2}$ thin films grown by similar methods \cite{shi_exchange_2004}. Further magnetic measurements of control samples can be found in the supplemental material of this manuscript \cite{VanHaren2023}.

\subsection{Electronic Transport Measurements}

\begin{figure}[h]
    \includegraphics{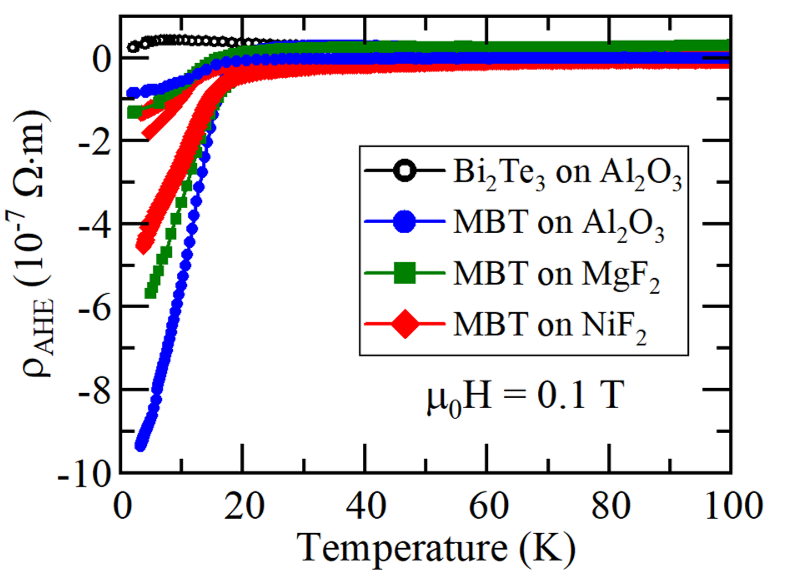}
    \caption{\label{RvT}Anomalous Hall resistivity as a function of temperature with $\mu_0 H = 0.1$~T, applied normal to the film surface, from a single run of all seven MBT films and a Bi$_2$Te$_3$ control film. Undoped Bi$_{2}$Te$_{3}$ film, shown in black, MBT grown on Al$_{2}$O$_{3}$, shown in blue, MBT grown on MgF$_{2}$, shown in green, and MBT grown on NiF$_{2}$, shown in red.}
    
\end{figure}

Figure \ref{RvT} shows the anomalous Hall resistivity as a function of temperature with an external magnetic field applied normal to the film surface of all seven MBT samples used in this study. From the onset of the AHE, the average transition temperature is calculated to be $T_C = 16.7 \pm 1.1$ K, with the error dominated by small variations of the transition temperature between samples. There is no evidence of an additional AHE at $ T = 78 $ K, the transition temperature of the NiF$_2$, in the MBT/NiF$_2$ bilayers. Studies of un-doped Bi$_2$Te$_3$/NiF$_2$ bilayers (not shown here) similarly showed no AHE. The magnitude of the AHE is believed to be driven by carrier density, with lower carrier densities correlating with a larger AHE \cite{chang_thin_2013}. Discussion of the carrier density is presented below.


\begin{figure*}
    \includegraphics{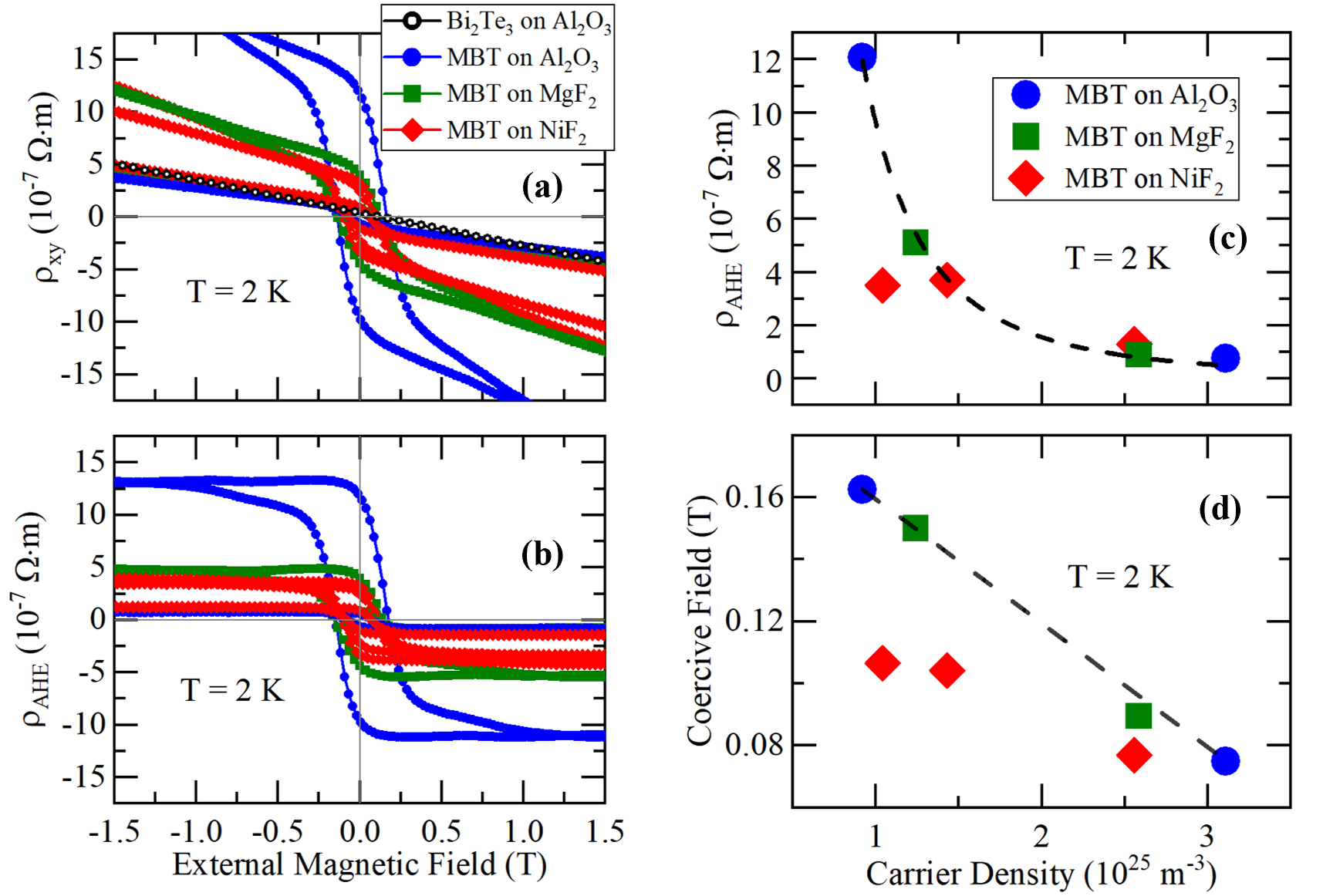}
    \caption{\label{MBT_Mag} (a) Hall resistivity as a function of applied field at $T = 2$~K, from a single run of all seven MBT films and a Bi$_2$Te$_3$ control film. Undoped Bi$_{2}$Te$_{3}$ film, shown in black, MBT grown on Al$_{2}$O$_{3}$, shown in blue, MBT grown on MgF$_{2}$, shown in green, and MBT grown on NiF$_{2}$, shown in red. (b) Anomalous Hall resistivity after subtraction of the ordinary Hall effect linear background. (c) AHE resistivity as a function of carrier density of all seven MBT films used in this study. (d) Coercive field values calculated from anomalous Hall resistivity measurements at $T = 2$ K of all seven MBT films. }
    
\end{figure*}

\begin{figure}
    \includegraphics{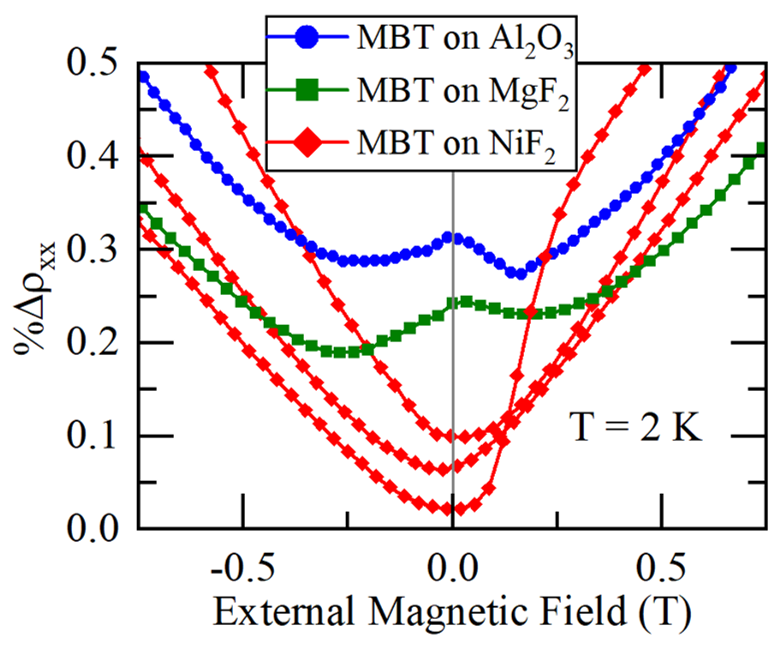}
    \caption{ \label{RxxvH} Longitudinal resistivity as a function of applied magnetic field of the lowest carrier density MBT films on Al$_2$O$_3$ and MgF$_2$, and all three MBT/NiF$_2$ bilayers.}
\end{figure}

The Hall resistivity of each sample was measured at $T = 2$~K as a function of applied magnetic field and the results are shown in Fig.~\ref{MBT_Mag}(a) for all seven MBT film samples and a control Bi$_2$Te$_3$ film. The carrier density of each sample was calculated from the ordinary Hall effect in regions where the AHE has saturated. The carriers in these films were found to be $n$-type, and the magnitudes of the carrier density for each sample are shown in Fig. \ref{MBT_Mag}(c,d). The variation in carrier density between samples is likely due to small differences in film thickness ($13.6 \pm 0.5$ nm) and Mn concentration ($7 \pm 1\%$), on average, between films. In these samples, thicker and lower Mn concentration films tend to have higher carrier densities. As seen in Fig. \ref{MBT_Mag}(c,d), the magnetic properties of the MBT films are correlated with the electron carrier density of the film. The saturation value of the AHE is observed to increase dramatically as the carrier density is decreased, as shown in Fig. \ref{MBT_Mag}(c). This effect could be due to increased magnetization in the MBT film, as the AHE is usually proportional to the magnetic moment orthogonal to the electronic current, but in TI systems such as this one, there is evidence of greatly enhanced AHE as the Fermi level approaches the bulk band gap, independent of the orthogonal magnetic moment \cite{chang_thin_2013}. In the films studied here, the lowest carrier density was calculated to be $0.92 \times 10^{25} ~ \text{m}^{-3}$, which suggests that the Fermi level of this film lies very close to the bottom of the bulk conduction band \cite{lee_ferromagnetism_2014}. In addition to the magnitude of the AHE, the coercive field of the MBT films also increased as the carrier density decreased, as shown in Fig. \ref{MBT_Mag}(d), but the coercive field of the MBT/NiF$_2$ bilayers was significantly lower than the MBT films on non-magnetic substrates. Magnetoresistance (MR) measurements of these films also showed ferromagnetic order. Figure \ref{RxxvH} shows the MR of the two lowest carrier density MBT films on non-magnetic substrates, and all three MBT/NiF$_2$ bilayers, at $T = 2 $ K. The small peaks in resistivity at zero field are due to weak localization (WL), a quantum mechanical effect of 2D electronic transport in systems with strong spin orbit coupling and local magnetic order. WL has been observed in ferromagnetic Cr-doped TI and is associated with the opening of the bulk band gap due to the spontaneous ferromagnetic moment breaking time reversal symmetry \cite{Hikami1980,kou_interplay_2013,liu_crossover_2012,liu_dimensional_2017,lu_competition_2011}. The WL seen here is relatively small, but it is entirely absent in the other samples with higher carrier density and in every MBT/NiF$_2$ bilayer, suggesting that the magnetic moment in these films is not sufficiently strong to open a large enough bulk band gap to induce WL. Longitudinal resistance measurements as a function of temperature can be found in the supplemental material of this manuscript \cite{VanHaren2023}.
There are two interesting conclusions that can be made from these results. The first is that the magnetization of the MBT films are enhanced as the Fermi level moves from the bulk conduction band towards the bulk band gap, which can be explained by surface state mediated magnetic ordering via the Ruderman-Kittel-Kasuya-Yosida (RKKY) interaction \cite{sessi_signatures_2014,checkelsky_dirac-fermion-mediated_2012,kou_interplay_2013}. This is unlike normal ferromagnetic metals, where the Stoner criterion dictates that a higher density of states at the Fermi level should increase the magnetic moment and the Curie temperature of the material~\cite{Stoner1938}. Therefore, our results demonstrate that the surface states in these MBT films play an important role in the spontaneous magnetization of this material that differentiates the magnetic ordering of this TI from ordinary metals. The second conclusion is that the NiF$_2$ interface appears to suppress the magnetization of the MBT films. In Fig.~\ref{MBT_Mag}(d), the coercive field of the MBT/NiF$_2$ bilayers is significantly lower than the coercive field of MBT films on non-magnetic substrates. WL is similarly suppressed in all MBT/NiF$_2$ bilayers, suggesting an insufficient magnetic moment to open a bulk band gap. A similar effect is seen in the magnetization measurements shown in Fig. \ref{MBT MvT}(a), where the magnetization at $ T = 4.5 $ K is lower in the MBT/NiF$_2$ bilayes than in the MBT films on Al$_2$O$_3$ or MgF$_2$. A possible mechanism responsible for the suppression of the ferromagnetic moment in MBT/NiF$_2$ bilayers is that the NiF$_2$ layer pins an interfacial layer of spins in the MBT to an in-plane orientation. It is known that ferromagnet/NiF$_2$ bilayers can exhibit exchange bias by a similar effect \cite{shi_exchange_2004}, and the MBT films studied here have a sufficiently weak magnetic anisotropy such that they can be, at least partially, rotated to an in plane orientation, as shown in Fig.~\ref{MBT MvT}(b).

\section{Conclusions}
In this study, we have shown how a spontaneous ferromagnetic moment below $T=16$~K can be induced in Mn$_{0.14}$Bi$_{1.86}$Te$_3$ thin films by doping Bi$_2$Te$_3$ with Mn atoms that randomly substitute into the Bi sites. We have presented evidence of successful MBT thin film growths on tetragonal crystal substrates, MgF$_2$ (110) and NiF$_2$ (110). Electronic transport and magnetic moment measurements show how the magnetization of the MBT films is enhanced as the Fermi level moves from the bulk conduction band towards the bulk band gap. When taken alongside previous works~\cite{checkelsky_dirac-fermion-mediated_2012,sessi_signatures_2014}, our study provides evidence that the electronic surface states of the TI play an important role in mediating ferromagnetic order in this material. This mechanism of magnetic ordering appears to be unique to the TI system, since the magnetization increases with decreasing volume carrier density. MBT/NiF$_2$ bilayers show evidence of a suppressed ferromagnetic moment along the [0001] direction, evidence of a proximity effect at the ferromagnetic MBT/antiferromagnetic NiF$_2$ interface. The magnetic behavior of these systems and their interactions represent an important piece of our understanding of these systems for potential device applications and the study of novel quantum phenomena. 

\begin{acknowledgments}
This work was supported in part by the Air Force MURI program, grant number FA9550-19-454963.
\end{acknowledgments}


\clearpage

\bibliography{MBTpaper}

\end{document}